\documentstyle[12pt]{article}

\def\be{\begin{equation}}
\def\ee{\end{equation}}
\def\ben{$$}
\def\een{$$}
\def\ba{\begin{array}{c}}
\def\ea{\end{array}}
\def\p{\partial}
\begin{document}

\titlepage
\vspace*{2cm}

 \begin{center}{\Large \bf
 ${\cal PT}-$supersymmetry\\
of singular harmonic oscillators
 }\end{center}

\vspace{10mm}

 \begin{center}
Miloslav Znojil

 \vspace{3mm}

\'{U}stav jadern\'e fyziky AV \v{C}R, 250 68 \v{R}e\v{z}, Czech
Republic\footnote{e-mail: znojil@ujf.cas.cz}

\end{center}

\vspace{5mm}

\section*{Abstract}

The Witten's supersymmetric interpretation of harmonic
oscillator in one dimension ($D = 1$) is generalized to any
$D>1$.  The emerging  centrifugal barriers are regularized via a
small imaginary shift of coordinates.  The resulting ${\cal PT}$
symmetrized supersymmetry $sl(1/1)$ interrelates the spiked
oscillator partners with different strengths of the barrier. The
innovated form of the creation and annihilation operators is
constructed.

\vspace{5mm}

PACS  11.30.Pb

\newpage

\section{Introduction}

Within the highly schematic formalism of the Witten's
supersymmetric quantum mechanics \cite{Witten}, many
difficult methodical problems related to the existence and
spontaneous breakdown of the supersymmetric multiplets of bosons
and fermions can be studied without any recourse to the less
essential subtleties of the relativistic quantum field theory.
The Witten's formalism recollects a years-long experience of
mathematical physicists with the ordinary differential
Schr\"{o}dinger equation in one dimension \cite{Darboux}. It
enables us to make a choice between the harmonic oscillator
$V(x) = x^2$ and other candidates.  The former, ``linear" model
is exceptional as it renders possible the construction of the
fermionic {\em plus} bosonic Fock space (cf.  review
\cite{Khare}, p.  285).

A significant drawback of the Witten's methodical
laboratory has been described by Jevicki and Rodrigues
\cite{Jevicki} as a partial or even complete failure of the
recipe whenever the (super)potentials happen to be
singular. We intend to
re-consider critically the real depth of the latter difficulty,
paying attention
to the more dimensional harmonic oscillators
which possess the ``solvable" centrifugal singularity.

Following the recent idea of Bender and Boettcher \cite{BB} and
working within their so called ${\cal PT}$ symmetric quantum
mechanics \cite{BBjmp,jmp}, we shall replace our
more-dimensional (and, admittedly, spiked) harmonic oscillator
Hamiltonians by the specific analytic continuation of their
radial components \cite{BG}.  In the other words, we shall work
with the complex, pseudo-Hermitian Hamiltonians \cite{whatis}
and assume for simplicity that their spectrum remains real and
discrete.

\section{Complexification of coordinates}


There hardly exists a textbook on quantum physics which would not
mention harmonic oscillator. The
shape of its potential can be used as a fairly realistic model in
atomic physics and quantum chemistry. Within nuclear theory, the
exact solvability of the $A-$body harmonic oscillator looks almost
like a miracle and proves particularly helpful. In field theory,
the simplicity of the annihilation and creation operators
 \be
 {\bf
 a}=\frac{\hat{q}+i\hat{p}}{\sqrt{2}} , \ \ \ \ \ \ \ \ \ \
 {\bf
 a}^\dagger=\frac{\hat{q}-i\hat{p}}{\sqrt{2}}
 \label{crea}
 \ee
enables us to understand free fields as ``local" excitations
$\Pi_\xi \,{\bf a}_\xi^\dagger\,|0\rangle$ of the vacuum.

In the recent studies of quantum fields  with broken parity
\cite{BM,kdu}, various Hamiltonians $H$ have been considered and
modified by the complex shift of coordinate,
 \be
 q \to r =r(x)=
 x-i\varepsilon, \ \ \ \ \ \ \ x \in (-\infty,\infty).
 \label{reg}
 \ee
The original symmetry with respect to the parity ${\cal P}$ was
broken and replaced by a significantly weaker ${\cal PT}$
symmetry $[H,{\cal PT}]=0$ where ${\cal T}$ mimicked the action
of the time-reversal of fields.
The Hermiticity of $H$ was lost.

Notably, once we work with a ``quasi-parity" quantum number
$\beta= \pm 1/2$, the usual spectrum
of the Hermitian harmonic oscillator
Hamiltonian
remains the same for all its shifted, non-Hermitian
descendants \cite{BB},
  \be
  E_{}^{}
  = E_{n}^{(\beta)} =4n+2\beta+2,
   \ \ \ \ \ \ \ \ \
    n = 0, 1,
  \ldots\
  \label{key}
  \ee
The preservation of the discrete, real and semi-bounded form of
these energies can be attributed to the ${\cal PT}$ symmetry of
the underlying Hamiltonian \cite{BBjmp}. Many exactly solvable
${\cal PT}$ symmetric models seem to support such a
hypothesis~\cite{kdu,others}.

Once we fix the shift $\varepsilon > 0$ in eq. (\ref{reg}), the
one-dimensional harmonic-oscillator
Schr\"{o}dinger equation can be replaced by
a family of its singular
generalizations
 \be
  -
 \frac{d^2}{dr^2} {\cal L}_{}^{}(r)
+
 \frac{\alpha^2-1/4}{r^2} {\cal L}_{}^{}(r)
+ r^2\,{\cal L}_{}^{}(r) = E_{}^{}
 {\cal L}_{}^{}(r)
 \label{SEa}
  \ee
with the same boundary conditions. The parameter
 \ben
\alpha=(D-2)/2 +\ell+ \omega > 0
 \een
varies with the spatial dimension $D$ and with the non-negative
integer angular momentum quantum number $\ell=0, 1, \ldots$.

Our previous energy formula (\ref{key}) remains valid and
continuous at $\alpha=1/2$ provided only that we put $\beta= \pm
\alpha$ \cite{hopt}.  Off the point $\alpha=1/2$ the spectrum of
our new equation $H^{(\alpha)}\psi=E\psi$ is richer
than in the Hermitian case. The analytic, confluent
hypergeometric origin of equation (\ref{SEa}) is to be seen in a
new perspective. The ordinary Sturm-Liouville theory must be
adapted to the new situation \cite{Sturm}. The norms have to be
replaced by the pseudo-norms \cite{whatime}.  At the same time,
our differential Schr\"{o}dinger equation (\ref{SEa}) stays {\em
safely regularized} at any~$\alpha > 0$.

In the generic case with $\alpha \neq 1/2$, our ${\cal PT}$
regularization ($\varepsilon \neq 0$) is only removable via the
limiting transition $\varepsilon \to 0$ accompanied by a halving
of the axis of coordinates.  This means that we have to replace
$r(x) = x - i\varepsilon$ by the radial and real $r\in (0,
\infty)$. We must cross out all the states with $\beta <
0$. The $\varepsilon \to 0$ return to the Hermitian cases $D=1$
or $D =3$ remains transparent.
A few details are mentioned in Table~1. The weird
difference between the linear and spherical Hermitian
oscillators can be attributed to the mere physically well
motivated difference in boundary conditions in the origin.

This initiated our study. Its decisive encouragement
came from the discovery of confluence of the different energy
levels at certain couplings $\alpha$.  This anomaly
remained unnoticed in ref. \cite{BG} and proves closely
connected to the ({\em unavoided}) crossing of the energy levels
at certain strengths of the barrier~\cite{hopt}. One can easily
verify that the latter puzzling phenomenon takes place at all
the integers $\alpha$, i.e., at all the even dimensions $D=2, 4,
\ldots$ for the vanishing residual $\omega=0$.

Such an observation is slightly discomforting. In the language
of linear algebra, the exceptional character of the integers
$\alpha$ can be intuitively explained via the occurrence of
Jordan blocks in the canonical form of the non-Hermitian
Hamiltonian matrix~\cite{Herbst}. In order to simplify the
discussion, we shall skip this point completely and, everywhere
in what follows, we shall assume that $\alpha \neq 0, 1, 2,
\ldots$.

\section{Supersymmetrization of pairs of Hamiltonians}

Let us first introduce a compact notation and recollect the
essence of the Witten's supersymmetric quantum mechanics. It
introduces the auxiliary ``conjugate" operators $A=\p_x+W$ and
$B=-\p_x+W$ in terms of the superpotential $W$. This
factorizes certain Hamiltonians in full accordance with the old
Schr\"{o}dinger's recipe \cite{Darboux}. For the two partner
operators
 \be
 H_{(L)}=\hat{p}^2+W^2-W'
 , \ \ \ \ \ \ \ \
 H_{(R)}=\hat{p}^2+W^2+W'
 \label{partner}
 \ee
it is easy to confirm that $H_{(L)}=B\cdot A$ while $H_{(R)}=A
\cdot B$. In the Witten's language, the related two by two
super-Hamiltonian and the two supercharges,
 \ben
 {
 \cal H}= \left [ \begin{array}{cc} H_{(L)}&0\\ 0&H_{(R)}
 \ea
 \right ]
, \ \ \ \ \ \
 {
 \cal Q}=\left [
 \begin{array}{cc} 0&0\\ A^{}&0
 \ea
 \right ],
 \ \ \ \ \ \
\tilde{\cal
Q}=\left [
 \begin{array}{cc}
0& B^{}
\\
0&0 \ea \right ]\
 \een
generate  a representation of the superalgebra sl(1/1). The
validity of the mixed anticommutation and commutation relations
 \ben
 \{ {\cal Q},\tilde{\cal Q}
\}={\cal H} , \ \ \ \ \ \ \{ {\cal Q},{\cal Q} \}= \{ \tilde{\cal
Q},\tilde{\cal Q} \}=0, \ \ \ \ \ \ \ \ [ {\cal H},{\cal Q} ]=[
{\cal H},\tilde{\cal Q} ]=0
  \een
mimics a supersymmetry between the bosonic and fermionic sectors
of the Hilbert space. As already mentioned, the most appealing
Fock-space picture of this type of the fermion -- boson
correspondence can be obtained in the ``linear model" using the
Hermitian one-dimensional harmonic oscillators. The scheme
collapses after one moves to any dimension $D > 1$ \{cf. the
counterexample (488) in review \cite{Khare}\}.

Our $\varepsilon \neq 0$ regularization of Schr\"{o}dinger
equations makes the spiked harmonic oscillators eligible for
supersymmetric treatment.
Let us return to our equation (\ref{SEa}) and to its well
known solvability in terms of Laguerre polynomials,
 \be
 {\cal L}_{ }^{}(r)
= {\cal L}_{n}^{(\beta)}(r) = \frac{n!}{\Gamma(n+\beta+1)}\cdot
 r^{\beta+1/2} \exp(-r^2/2) \cdot L_n^{(\beta)}(r^2)
 , \ \ \ \ \ \beta=\pm \alpha\ .
 \label{keyf}
  \ee
A few {\em algebraic} consequences follow from the analytic
${\cal PT}$ regularity at $\varepsilon \neq 0$.
We may pick up any real
parameter $\gamma \neq 0, \pm 1, \pm 2, \ldots$ and construct the
superpotential
 \be
 W_{}^{(\gamma)}(r) = - \frac{\p_r
 {\cal L}_{0}^{(\gamma)}(r)}{{\cal L}_{0}^{(\gamma)}(r)}=
r-\frac{\gamma+1/2}{r}\, .
 \label{N}
 \ee
The supersymmetric recipe gives the $\gamma-$numbered partner
Hamiltonians (\ref{partner}). After the appropriate insertions
they may be given the explicit and compact harmonic oscillator
form,
 \be
 {H}_{(L)}^{(\gamma)} = {H}_{}^{(\alpha)} -2\gamma-2,
 \ \ \ \ \ \
 {H}_{(R)}^{(\gamma)} = {H}_{}^{(\tilde{\alpha})} -2\gamma, \ \ \ \
 \ \ \ {\alpha}=|\gamma|, \ \ \ \
 \ \ \ \tilde{\alpha}=|\gamma+1|\ .
 \label{M}
 \ee
At all the non-integer $\gamma \neq 1/2$ the direct computations
reveal that the superscripted operators
 \be
 A^{(\gamma)}=\p_r+W^{(\gamma)}, \ \ \ \ \ \
 B^{(\gamma)}=-\p_r+W^{(\gamma)}, \ \ \ \ \ \ \gamma \neq 0, \pm
 1, \ldots
 \ee
act on our (normalized, spiked and ${\cal PT}-$symmetrized)
harmonic oscillator states in the transparent manner,
 \ben
 A^{(\gamma)}\, {\cal
 L}^{(\gamma)}_{n+1}{}=c_1(n,\gamma)\, {\cal
 L}^{(\gamma+1)}_{n}{}, \ \ \ \ \ \ \ \ \
\ c_1(n,\gamma)=-2\sqrt{n+1}; \een \ben B^{(\gamma)}\, {\cal
 L}^{(\gamma+1)}_{n}{}=c_2(n,\gamma)\, {\cal
 L}^{(\gamma)}_{n+1}{}, \ \ \ \ \ \ \ \ \
\ c_2(n,\gamma)=-2\sqrt{n+1}; \een
 \be A^{(\gamma)}\, {\cal
 L}^{(-\gamma)}_{n}{}=c_3(n,\gamma)\, {\cal
 L}^{(-\gamma-1)}_{n}{}, \ \ \ \ \ \ \ \ \
\ c_3(n,\gamma)=2\sqrt{n-\gamma};
 \label{rules}
 \ee
  \ben B^{(\gamma)}\, {\cal
 L}^{(-\gamma-1)}_{n}{}=c_4(n,\gamma)\, {\cal
 L}^{(-\gamma)}_{n}{} \ \ \ \ \ \ \ \ \
\ c_4(n,\gamma)=2\sqrt{n-\gamma}.
 \een
The former two rules were sufficient to define the well known
one-dimensional annihilation and creation at $\alpha = 1/2$. The
latter two lines have to be added in order to move us to any
$\alpha \neq 1/2$. Our operators mediate the supersymmetric
mapping and import an explicit $\gamma-$dependence in $c_3$ and
$c_4$.  Its definitely most embarrasing consequence is the
re-curring singularity at $\gamma=$integer. This reflects the
above-mentionend (unavoided) crossing of the energy levels.

\section{Transitions between the different $\alpha$}

We observe that each of the two partner Hamiltonians generates the
two quasi-parity subsets of the equidistant energy values. At each
$ n = 0, 1, \ldots$, the quadruplet of energies
 \be
 {E}_{(L)}^{(+)} = 4n, \ \ \ \ \
 {E}_{(L)}^{(-)} =
 4n-4\gamma, \ \ \ \ \ \
 {E}_{(R)}^{(+)} = 4n+4, \ \ \ \ \
 {E}_{(R)}^{(-)} = 4n-4\gamma
 \label{K}
 \ee
corresponds to the respective wave functions
 \ben
   {{\cal L}}_{n}^{(\gamma)}, \ \ \ \ \
 {{\cal L}}_{n}^{(-\gamma)}, \ \ \ \ \
 {{\cal L}}_{n}^{(\gamma+1)}, \ \ \ \
 {{\cal L}}_{n}^{(-\gamma-1)}\ .
 \een
The ordering of these states depends on the sign of the parameter
$\gamma$ in a way summarized in Table~2. In this scheme, a mere
shift of the energy spectrum (i.e., a coincidence of the
Hamiltonians with $\alpha$ and $ \tilde{\alpha}$) is, obviously,
exceptional. The related parameter $\gamma_e$ is given by the
algebraic equation $|\gamma_e|=|\gamma_e+1|$ with the unique
solution $\gamma_e = -1/2$. In the Hermitian limit $\varepsilon
\to 0$ we reproduce the one-dimensional example. Its supersymmetry
is unbroken.

A few nontrivial $\alpha \neq 1/2$ examples of our generalized
supersymmetric partnership are displayed in the pairs of the
neighboring columns in Table~3. Its part (a) samples the two-way
correspondence between the two different Hamiltonians
$H^{(1/2)}$ and $H^{(3/2)}$. The $\gamma=-3/2$ ${\cal PT}$
supersymmetry between ${H_{(L)}=H^{(3/2)}+1}$ and
$H_{(R)}=H^{(1/2)}+3$ is followed by the  $\gamma=1/2$
correspondence between the doublet
${H_{(\tilde{L})}}=H^{(1/2)}-3$ and
${H_{(\tilde{R})}=H^{(3/2)}-1}$. As a net result we obtain the
appropriate generalization of the creation/annihilation pattern
for the $p-$wave column of Table~1.

At $\gamma=-1/2$ we encounter the ``degenerate" (and, in the
present context, utterly exceptional) textbook $\gamma=-1/2$
pattern
 \ben
 {\bf a}\cdot  {\cal L}_{n-1}^{(1/2)}(x)
 = \sqrt{2n-1}\,
   {\cal L}_{n-1}^{(-1/2)}(x) ,
 \ \ \ \ \ \ \ \ \
 {\bf a}\cdot  {\cal L}_{n}^{(-1/2)}(x)
 = - \sqrt{2n}\,  {\cal L}_{n-1}^{(1/2)}(x)
 \een
 \ben
 {\bf a}^\dagger\,
{\cal L}_{n-1}^{(-1/2)}(x)
 = \sqrt{2n-1}\,  {\cal L}_{n-1}^{(1/2)}(x) , \ \ \ \ \
\ \ \ \
 {\bf a}^\dagger\, {\cal L}_{n-1}^{(1/2)}(x)
 = - \sqrt{2n}\,  {\cal L}_{n}^{(-1/2)}(x)
 \een
where ${\bf a} \sim A^{(-1/2)}$ and ${\bf a}^\dagger \sim
B^{(-1/2)}$. In a slightly re-ordered form, Table 3 (a) offers
an alternative. Via the non-Hermitian detour and limit
$\varepsilon \to 0$,  another explicit annihilation pattern is
obtained for the same $s-$wave oscillator.  The new two-step
mapping starts from the Hamiltonian ${H_{(L)}=H^{(1/2)}-3}$ and
gives, firstly, its ${\cal PT}$ symmetrically regularized
non-Hermitian supersymmetric partner $H_{(R)}=H^{(3/2)}-1$ [cf
the last two columns in Table 3 (a)].  In the subsequent step,
the similar partnership of the re-shifted $
H_{(\tilde{L})}=H^{(3/2)}+1$ returns us to the original
${H_{(\tilde{R})}=H^{(1/2)}+3}$ [note that the shift differs
from the one used in the first two columns in Table 3 (a)].

\section{Creation and annihilation operators}

At any $\gamma \neq -1/2$, our new supersymmetric pattern differs
quite significantly from the Hermitian one. Firstly, in a fairly
unusual way, the states can easily appear at a negative value of
the energy [cf. the $E_{(L)0}^{(-)}=E_{(R)0}^{(-)}=-2$ right-hand
side sample in Table~3 (a) etc].  Secondly, in a way reflecting
the same possibility, the usual ``unmatching" (i.e., absence of
the $_{(R)}-$subscripted partner) at the vanishing energy
$E_{(L)}^{(\beta)}=0$ can occur for the excited states [cf., e.g.,
${\cal L}^{(3/2)}_{0}{}\rightarrow 0$ in Table 3 (b); the
existence of such an anomalous ``excited vacuum" is particularly
important for the non-equidistant spectra]. Thirdly, in a sharp
contrast to the Hermitian models, the lowest state exists
(i.e., remains normalizable) in both the spectra of $H_{(L)}$ and
$H_{(R)}$ [cf., e.g., the $\gamma=3/2$ and $E=-6$ example ${\cal
L}^{(-3/2)}_{0}{}\rightarrow {\cal L}^{(-5/2)}_{0}$ in Table 3
(b)].

What can we expect from moving to the larger
semi-integers $\alpha = |\gamma| $? Just a strengthening of the
tendencies which were revealed in Table 3.  Their features can be
simply extrapolated. Thus, the large$-\gamma$ modifications of
Table 3 will contain more lines at the bottom. For example, the
$\gamma=5/2$ supersymmetry between ${H_{(L)}=H^{(5/2)}-7}$ and
${H_{(R)}=H^{(7/2)}-5}$ will induce a ground-state mapping ${\cal
L}^{(-5/2)}_{0}{} {\stackrel{A^{(5/2)} }{\longrightarrow}} {\cal
L}^{(-7/2)}_{0}{}$. It appears at $ E_{(L/R)}=-10$, witnessing
just a continuing downward shift of the levels with the negative
and decreasing superscripts.

Very similar conclusions can be drawn from the results generated
at any real superscript value $\gamma$.
We are near a climax of our study. A nice and transparent
structure emerges form all the above ${\cal PT}$ supersymmetric
assignments. Having spotted the difference between the regular and
quasi-singular supersymmetries with $\alpha=1/2$ and $\alpha \neq
1/2$, respectively, we are ready to study more deeply the
non-equidistant spectra corresponding to the non-integer values of
$2\alpha$.

Unless we reach the points of
degeneracy $\alpha =$integer, all our formulae remain
applicable. The annihilation operators and their creation
partners acquire the factorized, second-order differential form
\ben
A^{(-\gamma-1)}
\cdot
A^{(\gamma)}
=
A^{(\gamma-1)}
\cdot
A^{(-\gamma)}
=
{\bf A}(\alpha),
 \een
 \ben
B^{(-\gamma)}
\cdot
B^{(\gamma-1)}=
B^{(\gamma)}
\cdot
B^{(-\gamma-1)}=
{\bf A}^\dagger (\alpha).
 \een
At any $ \alpha \neq 0, 1, 2, \ldots$, they enable us to move
along the spectrum of any harmonic oscillator Hamiltonian
$H^{(\alpha)}$. We get
\ben
 {\bf A}(\alpha) \cdot
 {\cal L}^{(\beta)}_{n+1}{}=c_5(n,\beta)\,
 {\cal L}^{(\beta)}_{n}{},
 \een
 \ben
{\bf A}^\dagger(\alpha) \cdot
 {\cal L}^{(\beta)}_{n}{}=c_5(n,\beta)\,
 {\cal L}^{(\beta)}_{n+1}{},
 \een
 \ben
c_5(n,\beta)=-4\sqrt{(n+1)(n+\beta+1)},
\ \ \ \ \ \ \ \ \beta=\pm \alpha.
 \een
This action is elementary and transparent.

\section{Summary}

We achieved a unified description of the spiked harmonic
oscillators $H^{(\alpha)}$ within the ${\cal PT}$ symetric
framework.

\begin{itemize}

\item
The general ${\cal PT}$ supersymmetric partnership has been
shown mediated by the ``shape-invariance" operators
$A^{(\gamma)}$ and $B^{(\gamma)}$.

\item
At any non-integer $\alpha>0$ the role of the general creation
and annihilation operators {\em for a given, single} Hamiltonian
$H^{(\alpha)}$ has been shown played by their $\alpha-$dependent
and $\beta-$preserving products ${\bf A}^\dagger(\alpha)$ and
${\bf A}(\alpha)$, respectively.

\end{itemize}

 \noindent
The only case where, up to a constant shift, the ${\cal PT}$
supersymmetric partners coincide corresponds to the case where the
poles in $A^{(\gamma)}$ and $B^{(\gamma)}$ vanish.  This is the
only case tractable {\em without} the use of the ${\cal PT}$
regularization.  In this sense, the traditional creation and
annihilation using ${\bf a}^\dagger=B{(-1/2)}/\sqrt{2}$ and
${\bf a}=A{(-1/2)}/\sqrt{2}$ is slightly misleading since these
operators {\em change} the quasi-parity $\beta $ to $-\beta$.
Such a correpsondence is not transferrable to any
non-equidistant spectrum with $\alpha \neq -1/2$.

Our ``natural" operators of creation ${\bf A}^\dagger(\alpha)$
and annihilation ${\bf A}(\alpha)$ are smooth near $\alpha =
1/2$.  Their marginal (though practically relevant) merit lies
in their reducibility to their regular (and hence, of course,
state-dependent) first-order differential representation
 \ben
 {\bf A}(\alpha) \cdot
 {\cal L}^{(\beta)}_{n}{}=
  (2r\p_r+2\,r^2-4n-2\beta-1)\cdot
 {\cal L}^{(\beta)}_{n}{},
 \een
 \ben
 {\bf A}^\dagger(\alpha) \cdot
 {\cal L}^{(\beta)}_{n}{}=
  (-2r\p_r+2\,r^2-4n-2\beta-3)\cdot
 {\cal L}^{(\beta)}_{n}{}.
 \een
Let us notice that the change of variables $r \to y$ giving a
simpler differentiation $2r\p_r \to \p_y$ reads $r=\exp 2y$ and
would result in the so called Morse form of the
Hamiltonian. This could, in principle, indicate
that the Morse potentials with ${\cal PT}$ symmetry~\cite{Morse}
would deserve more attention.

\section*{Acknowledgement}

Work supported by the grant Nr. A 1048004 of GA AS CR.

\newpage

\section*{Tables}


Table 1.

 \noindent
A ``disappearance" of wave functions in the Hermitian limit
 $\varepsilon \to 0$ at $D = 3$.

 $$
\begin{array}{||rcl||rcl|rcl||c||}
\hline \hline
  \multicolumn{9}{||c||}{{\rm harmonic\ oscillator}}&
  {\rm energy}\\
  \hline \hline
 \multicolumn{3}{||c||}{{\rm linear}\ (D=1)}&
 \multicolumn{6}{c||}{{\rm spherical\ }(D=3)}&
 {\rm }\\
 \multicolumn{3}{||c||}{{\rm }}&
\multicolumn{3}{c}{{\rm }s-{\rm wave}} &\multicolumn{3}{c||}{{\rm
}p-{\rm wave}}& {\rm }\\
 \hline
&&\vdots&&&\vdots&&&\vdots&\vdots
\\ |4\rangle&=&{\cal L}_2^{(-1/2)}& && - &
|1\rangle&=&\sqrt{2}{\cal L}_2^{(3/2)}&9\\ |3\rangle&=&{\cal
L}_1^{(1/2)}& |1\rangle&=&\sqrt{2}{\cal L}_1^{(1/2)}& & &-&7\\
|2\rangle&=&{\cal L}_1^{(-1/2)}& & &-& |0\rangle&=&\sqrt{2}{\cal
L}_1^{(3/2)}&5\\ |1\rangle&=&{\cal L}_0^{(1/2)}&
|0\rangle&=&\sqrt{2}{\cal L}_0^{(1/2)}&& &- &3\\ |0\rangle&=&{\cal
L}_0^{(-1/2)}& & &- &&&- &1\\ &&&&&&&&&\\ \hline \hline \ea $$

\newpage

Table 2.

 \noindent
Supersymmetry of harmonic oscillators at non-integer $\alpha =
|\gamma|$. Superpotential (\ref{N}) gives the partner Hamiltonians
(\ref{partner}) and quadruplets of energies $E_a\leq E_b\leq
E_c\leq E_d$ at each $n=0, 1, \ldots$.

 $$
 \begin{array}{||c||c|c|c||}
 \hline \hline
  {\rm range\ of\ }\alpha
  &(0, \infty)& (0,1)&
  (1,\infty)\\
  \hline
  \hline
 {\rm parameters}&&&\\
 \tilde{\alpha}=|\gamma+1|&\alpha+1&1-\alpha&\alpha-1\\
   {\rm  }\gamma&\alpha&-\alpha&-\alpha\\
  &&&\\
  \hline
  \hline
 {\rm Hamiltonians}&&&\\
  H_{(L)}^{(\gamma)}&H^{(\alpha)} -2\tilde{\alpha}&H^{(\alpha)}
  -2\tilde{\alpha}&H^{(\alpha)}+
  2\tilde{\alpha}\\
  H_{(R)}^{(\gamma)}&H^{(\tilde{\alpha})}-2\alpha&H^{(\tilde{\alpha})}
  +2\alpha&H^{(\tilde{\alpha})}+2\alpha\\
  &&&\\
 \hline
  \hline
 {\rm energies}&&&\\
    E_d=E_{(R)}\left [ {\rm of\ }
  {\cal L}_{n}^{(\tilde{\alpha})}
  \right ]&4n+4&4n+4&4n+4\alpha\\
    E_c=E_{(L)}\left [ {\rm of\ }
  {\cal L}_{n}^{(\alpha)}
  \right ]&4n&4n+4\alpha&4n+4\alpha\\
    E_b=E_{(R)}\left [ {\rm of\ }
  {\cal L}_{n}^{(-\tilde{\alpha})}
  \right ]&4n-4\alpha&4n+4\alpha&4n+4\\
  E_a=E_{(L)}\left [ {\rm of\ }
  {\cal L}_{n}^{(-\alpha)}
  \right ]&4n-4\alpha&4n&4n\\
  &&&\\
  \hline
 \hline \ea $$

\newpage

 Table 3.

\noindent Supersymetry for the singular Hamiltonian
$H^{(3/2)}=p^2+(x-i\varepsilon)^2+2/(x-i\varepsilon)^2$.

\noindent (a) $\gamma = -3/2$ and  $-\gamma-1 = 1/2$.

$$
\begin{array}{||c|ccccc|c||}
\hline \hline E_{(L/R)} &|n_{(L)}\rangle& {\stackrel{A^{(-3/2)}
}{\longrightarrow}} &|n_{(R)}\rangle= |n_{(\tilde{L})}\rangle&
{\stackrel{A^{(1/2)} }{\longrightarrow}} &|n_{(\tilde{R})}\rangle&
E_{(\tilde{L}/\tilde{R})}
\\
 \hline
\hline \vdots&\vdots&&\vdots&&\vdots&\vdots
\\
 14&{\cal L}^{(3/2)}_{2}{}&\rightarrow&{\cal L}^{(1/2)}_{2}{}&
\rightarrow&{\cal L}^{(3/2)}_{1}{}&8
 \\
 12&{\cal L}^{(-3/2)}_{3}{}&\rightarrow&{\cal L}^{(-1/2)}_{2}{}&
\rightarrow&{\cal L}^{(-3/2)}_{2}{}&6
 \\
 10&{\cal L}^{(3/2)}_{1}{}&\rightarrow&{\cal L}^{(1/2)}_{1}{}&
\rightarrow&{\cal L}^{(3/2)}_{0}{}&4
 \\
 8&{\cal L}^{(-3/2)}_{2}{}&\rightarrow&{\cal L}^{(-1/2)}_{1}{}&
\rightarrow&{\cal L}^{(-3/2)}_{1}{}&2
 \\
 6&{\cal L}^{(3/2)}_{0}{}&\rightarrow&{\cal L}^{(1/2)}_{0}{}
&\rightarrow&0&0
 \\
 4&{\cal L}^{(-3/2)}_{1}{}&\rightarrow&{\cal L}^{(-1/2)}_{0}{}&
 \rightarrow&{\cal L}^{(-3/2)}_{0}{}&-2
 \\
2&-&&-&&-&-4\\
 0&{\cal L}^{(-3/2)}_{0}{}&\rightarrow&{0}{}&
 \rightarrow&-&-6
 \\
 \hline \hline
\ea $$

\noindent (b) $\gamma = 3/2$ and  $-\gamma-1 = -5/2$.

$$
\begin{array}{||c|ccccc|c||}
\hline \hline E_{(L/R)} &|n_{(L)}\rangle& {\stackrel{A^{(3/2)}
}{\longrightarrow}} &|n_{(R)}\rangle= |n_{(\tilde{L})}\rangle&
{\stackrel{A^{(-5/2)} }{\longrightarrow}}
&|n_{(\tilde{R})}\rangle& E_{(\tilde{L}/\tilde{R})}
\\
 \hline
\hline \vdots&\vdots&&\vdots&&\vdots&\vdots
\\
 8&{\cal L}^{(3/2)}_{2}{}&\rightarrow&{\cal L}^{(5/2)}_{1}{}&
\rightarrow&{\cal L}^{(3/2)}_{1}{}&14
 \\
 6&{\cal L}^{(-3/2)}_{3}{}&\rightarrow&{\cal L}^{(-5/2)}_{3}{}&
\rightarrow&{\cal L}^{(-3/2)}_{2}{}&12
 \\
 4&{\cal L}^{(3/2)}_{1}{}&\rightarrow&{\cal L}^{(5/2)}_{0}{}&
\rightarrow&{\cal L}^{(3/2)}_{0}{}&10
 \\
 2&{\cal L}^{(-3/2)}_{2}{}&\rightarrow&{\cal L}^{(-5/2)}_{2}{}&
\rightarrow&{\cal L}^{(-3/2)}_{1}{}&8
 \\
 0&{\cal L}^{(3/2)}_{0}{}&\rightarrow&0&\rightarrow&-&6
 \\
 -2&{\cal L}^{(-3/2)}_{1}{}&\rightarrow&{\cal L}^{(-5/2)}_{1}{}&
 \rightarrow&{\cal L}^{(-3/2)}_{0}{}&4
 \\
-4&-&&-&&-&2\\
 -6&{\cal L}^{(-3/2)}_{0}{}&\rightarrow&{\cal L}^{(-5/2)}_{0}{}&
 \rightarrow&0&0
 \\
 \hline \hline
\ea $$


\newpage

\begin{thebibliography}{00}

\bibitem{Witten}
Witten E 1981 Nucl. Phys. B 188 513;

Witten E 1982 Nucl. Phys. B 202 253

\bibitem{Darboux}
Liouville J 1837 J. Math. Pures Appl 1 16;

Infeld L and Hull T E 1951
               Rev. Mod. Phys. 23 21

\bibitem{Khare}
Cooper F,  Khare A and Sukhatme U 1995 Phys. Rep. 251 267

\bibitem{Jevicki}
Jevicki A and Rodrigues J 1984 Phys. Lett. B 146 55

\bibitem{BB}
Bender C M and Boettcher S 1998 Phys. Rev. Lett.  24  5243

\bibitem{BBjmp}
Bender C M, Boettcher S and Meisinger P N 1999 J. Math. Phys. 40 2201

\bibitem{jmp}
Caliceti E, Graffi S and Maioli M 1980 Commun. Math. Phys. 75 51;

Bender C M and Turbiner A 1993 Phys. Lett.  A 173  442;

Cannata F, Junker G and Trost J 1998 Phys. Lett.  A 246 219;

Delabaere E and Pham F 1998 Phys. Lett. A 250 25 and 29;

Andrianov A A, Ioffe M V, Cannata F and Dedonder J P 1999 Int.
J. Mod. Phys. A 14 2675;

Mezincescu G A 2000 J. Phys. A: Math. Gen. 33 4911;

Delabaere E and Trinh D T 2000 J. Phys. A: Math. Gen. 33
8771;

Znojil M and L\'{e}vai G 2000 Phys. Lett. A 271 327;

Znojil M and Tater M 2001 J. Phys. A: Math. Gen. 34 1793

\bibitem{BG}
Buslaev V and Grecchi V 1993  J. Phys. A: Math. Gen. 26 5541

\bibitem{whatis}
Fernandez F M, Guardiola R, Ros J and Znojil M 1998
J. Phys. A: Math. Gen. 31 10105;

Znojil M 2001 ``What is PT symmetry" (arXiv: quant-ph/0103054)

\bibitem{BM}
Bender C M and Milton K A 1997 Phys. Rev. D 55 R3255;

Bender C M and Milton K A 1998 Phys. Rev. D 57 3595;

Bender C M and Milton K A 1999 J. Phys. A: Math. Gen. 32 L87

\bibitem{kdu}
Fernandez F M, Guardiola R, Ros J and Znojil M 1999
J. Phys. A: Math. Gen. 32 3105;

Znojil M 2000 J. Phys. A: Math. Gen. 33 L61 and 4203 and 4561 and 6825;

L\'{e}vai G and Znojil M 2000 J. Phys. A: Math. Gen.  33 7165

\bibitem{others}
Bender C M and Boettcher S 1998 J. Phys. A: Math. Gen. 31 L273;

Znojil M 1999 J. Phys. A: Math. Gen. 32 4563;

Bagchi B and Roychoudhury R 2000 J. Phys. A: Math. Gen. 33 L1;

Bagchi B, Cannata F and  Quesne C 2000 Phys. Lett. A 269  79;

Znojil M, Cannata F, Bagchi B and Roychoudhury R 2000
         Phys. Lett. B 483 284;

Bagchi B and Quesne C 2000 Phys. Lett. A 273 285;

Khare A and Mandal B P 2000 Phys. Lett. A 272  53;

Bagchi B, Mallik S and Quesne C 2001 Int. J. Mod. Phys. A, to
appear (arXiv: quant-ph/0102093)

\bibitem{hopt}
Znojil M 1999 Phys. Lett. A 259  220

\bibitem{Sturm}
Hille E 1969 Lectures on Ordinary Differential Equations
(Reading, MA: Addison-Wesley);

Bender C M, Boettcher S and Savage V M 2000 J. Math. Phys. 41 6381

\bibitem{whatime}
Znojil M 2001 ``Conservation of pseudonorm in PT symmetric
quantum mechanics" (arXiv: math-ph/0104012)

\bibitem{Herbst}
Herbst I 2000 private communication

\bibitem{Morse}
Znojil M 1999 Phys. Lett. A 264 108

\end{thebibliography}
\end{document}